\documentclass[aps,twocolumn,amsmath,amssymb,nofootinbib,superscriptaddress,floatfix,reprint,longbibliography]{revtex4-1}
\usepackage[dvips]{graphicx}
\usepackage{latexsym}
\usepackage{amsmath}
\usepackage{amsfonts}
\usepackage{amssymb}
\usepackage{bm}
\usepackage{color}
\usepackage{txfonts}
\usepackage{float}
\usepackage{makecell}
\usepackage{url}
\usepackage[colorlinks=true, urlcolor=blue, linkcolor=blue, citecolor=blue, pdftex]{hyperref}
\usepackage{ulem}
\usepackage{physics}
\usepackage{hhline}
\begin{document}
	\newcommand{\fig}[2]{\includegraphics[width=#1]{#2}}
	\newcommand{\pprl}{Phys. Rev. Lett. \ }
	\newcommand{\pprb}{Phys. Rev. {B}}

\title {Recent progress in nickelate superconductors}

\author{Yuxin Wang}
\affiliation{Beijing National Laboratory for Condensed Matter Physics and Institute of Physics,
	Chinese Academy of Sciences, Beijing 100190, China}
\affiliation{School of Physical Sciences, University of Chinese Academy of Sciences, Beijing 100190, China}

\author{Kun Jiang}
\email{jiangkun@iphy.ac.cn}
\affiliation{Beijing National Laboratory for Condensed Matter Physics and Institute of Physics,
	Chinese Academy of Sciences, Beijing 100190, China}
\affiliation{School of Physical Sciences, University of Chinese Academy of Sciences, Beijing 100190, China}

\author{Jianjun Ying}
\affiliation{Hefei National Laboratory for Physical Sciences at the Microscale, University of Science and
	Technology of China, Hefei, Anhui 230026, China}
\affiliation{CAS Key Laboratory of Strongly-coupled Quantum Matter Physics, Department of Physics, 
	University of Science and Technology of China, Hefei, Anhui 230026, China}

\author{Tao Wu}
\email{wutao@ustc.edu.cn}
\affiliation{Hefei National Laboratory for Physical Sciences at the Microscale, University of Science and
	Technology of China, Hefei, Anhui 230026, China}
\affiliation{CAS Key Laboratory of Strongly-coupled Quantum Matter Physics, Department of Physics, 
	University of Science and Technology of China, Hefei, Anhui 230026, China}

\author{Jinguang Cheng}
\affiliation{Beijing National Laboratory for Condensed Matter Physics and Institute of Physics,
	Chinese Academy of Sciences, Beijing 100190, China}
\affiliation{School of Physical Sciences, University of Chinese Academy of Sciences, Beijing 100190, China}

\author{Jiangping Hu}
\email{jphu@iphy.ac.cn}
\affiliation{Beijing National Laboratory for Condensed Matter Physics and Institute of Physics,
	Chinese Academy of Sciences, Beijing 100190, China}
\affiliation{Kavli Institute of Theoretical Sciences, University of Chinese Academy of Sciences,
	Beijing, 100190, China}
 \affiliation{New Cornerstone Science Laboratory, 
	Beijing, 100190, China}

\author{Xianhui Chen}
\email{chenxh@ustc.edu.cn}
\affiliation{Hefei National Laboratory for Physical Sciences at the Microscale, University of Science and
	Technology of China, Hefei, Anhui 230026, China}
\affiliation{CAS Key Laboratory of Strongly-coupled Quantum Matter Physics, Department of Physics, 
	University of Science and Technology of China, Hefei, Anhui 230026, China}

\date{\today}

\begin{abstract}
The discovery of superconductivity in nickelate compounds has opened new avenues in the study of high-temperature superconductors. Here we provide a comprehensive overview of recent progress in the field, including all different nickelate systems, reduced-Ruddlesden-Popper-type infinite layer LaNiO$_2$, Ruddlesden-Popper-type bilayer La$_3$Ni$_2$O$_7$ and trilayer La$_4$Ni$_3$O$_{10}$. We begin by introducing the superconducting properties of the hole-doped LaNiO$_2$ system, which marked the starting point for nickelate superconductivity. We then turn to the bilayer La$_3$Ni$_2$O$_7$ system, discussing both its high-pressure and thin-film superconducting phases. This is followed by an examination of the trilayer La$_4$Ni$_3$O$_{10}$ system and other related multilayer nickelates. Throughout the review, we highlight emerging trends, key challenges, and open questions. We conclude by addressing current limitations in materials synthesis and characterization, and future directions that may help uncover the mechanisms driving superconductivity in these complex oxide systems.
\end{abstract}
\maketitle

\section{Introduction}
The discovery of superconductivity in copper oxides in 1986 opened a new frontier in the quest for high-temperature—and potentially even room-temperature—superconductors \cite{labacuo,doping_mott,keimer_review}. These materials marked not only a major step toward real-world applications but also unveiled a rich landscape of exotic quantum phenomena. In high-temperature (high-$T_c$) superconductors, electrons interact strongly in ways that defy the conventional Bardeen-Cooper-Schrieffer (BCS) theory, revealing complex behaviors driven by electronic correlations. Their ability to sustain superconductivity within layered structures containing intertwined magnetic and electronic orders sparked a wave of research into unconventional pairing mechanisms. The discovery of iron-based superconductors in 2008 further broadened the family of high-$T_c$ materials \cite{iron1,iron2,iron_review}. Yet, despite decades of intense investigation, the underlying mechanisms that drive high-temperature superconductivity remain one of the most profound and enduring mysteries in modern condensed matter physics.
One promising approach to better understand unconventional pairing is to explore and identify more high-$T_c$ superconductors.



Nickel, as the nearest neighbor of copper on the periodic table of elements, has long been considered a promising candidate for hosting high-temperature superconductivity since early 1990s. Notably, Maurice Rice and collaborators proposed that superconductivity could emerge in doped LaNiO$_2$, given that Ni$^+$ shares the same 3$d^9$ valence electron configuration as Cu$^{2+}$ in cuprate superconductors \cite{rice_PhysRevB.59.7901,pickett_PhysRevB.70.165109}. After more than three decades of exploration, this prediction was finally realized with the discovery of superconductivity in nickelates \cite{lidanfeng}, marking the beginning of the ``nickel age" of superconductivity \cite{norman2020,pickett-2021}. Recently, the family of nickelate superconductors has grown rapidly, now including compounds such as La$_3$Ni$_2$O$_7$ \cite{meng_wang,chengjg_nature}, La$_4$Ni$_3$O$_{10}$ \cite{zhaojun,Qiyp_PhysRevX.15.021005,haihu_cpl}, and other related materials \cite{mundy-np,la5ni3o11}. These discoveries have introduced a new and exciting chapter in the study of high-$T_c$ superconductors. The purpose of this paper is to review the recent advances in nickelate superconductors and to provide an accessible introduction to this rapidly evolving field.

\begin{figure*}
	\begin{center}
		\fig{7.0in}{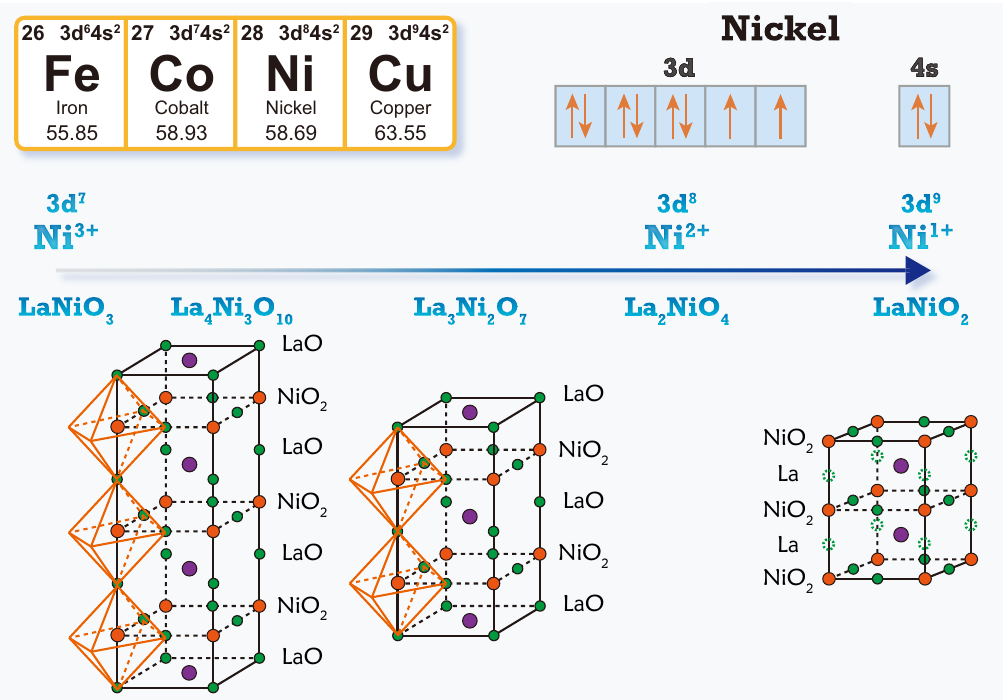}\caption{\textbf{Structures, Ionic Charges, and Valence States of Nickelates.} Ni atom has electron configuration [Ar] 3$d^8$ 4$s^2$ or [Ar] 3$d^9$ 4$s^1$. In nickel oxides, the Ni ionic charge typically ranges from $+1$ to $+3$ and the most common oxidation state is Ni$^{2+}$. The RP nickelates phases La$_{n+1}$Ni$_n$O$_{3n+1}$ have ionic charges between Ni$^{2+}$ and Ni$^{3+}$. The bilayer compound La$_3$Ni$_2$O$_7$ and the trilayer La$_4$Ni$_3$O$_{10}$ have recently been identified as superconductors. A reduced RP phase, LaNiO$_2$, can be obtained from the RP LaNiO$_3$ through chemical reduction. These removed oxygen sites are indicated by dashed green circles in LaNiO$_2$.
        Hole-doped LaNiO$_2$ is the first nickelate superconductor.
			\label{fig1}}
	\end{center}
	\vskip-0.5cm
\end{figure*}

From the valence electron perspective, the electron distribution of free Ni atom can be formally written as [Ar]3$d^8$4$s^2$ or [Ar]3$d^9$4$s^1$. The most common oxidation state of Nickel is Ni$^{2+}$ while the ionic charge of Ni oxides ranges from $1+$ to $3+$, as plotted in Fig.\ref{fig1}. As mentioned above, Ni$^{1+}$ has 3$d^9$ valence electrons while Ni$^{2+}$, Ni$^{3+}$ belong to 3$d^8$ or 3$d^7$ respectively. We want to add one caveat that for ionic charge larger than 2 state Ni$^{>2+}$, the extra holes are always doped to oxygen as Zhang-Rice singlets in cuprates \cite{zhang_rice_PhysRevB.37.3759}, where the oxygen $p$ orbital is also important \cite{sawatzky_PhysRevB.45.1612}.

From a structural perspective, nickelate superconductors belong to the well-known Ruddlesden-Popper (RP) family of perovskite structures \cite{RP-phase}. The general chemical formula for an RP phase is A$_{n+1}$Ni$_n$O$_{3n+1}$, where $A$ is a cation and $n$ denotes the number of perovskite layers. For clarity, we consider the case with lanthanum (La) as the $A$-site ion, as illustrated in Fig. \ref{fig1}. RP phases consist of multiple perovskite-like LaNiO$_3$ layers interleaved by single rock-salt-type LaO layers. For $n = 1$, the resulting structure is La$_2$NiO$_4$, where nickel exists in a 2+ oxidation state. Increasing to $n = 2$ yields the bilayer compound La$_3$Ni$_2$O$_7$, and $n = 3$ gives the trilayer La$_4$Ni$_3$O$_{10}$. The corresponding average nickel oxidation states are 2.5+ and 2.67+, respectively. In the limit $n \rightarrow \infty $, the structure converges to the compound LaNiO$_3$, in which nickel has a 3+ oxidation state. Thus, the valence state of nickel across the RP nickelates spans from Ni$^{2+}$ and Ni$^{3+}$. To achieve a Ni $d^9$ electronic configuration, analogous to that of Cu$^{2+}$ in cuprates, one can remove an oxygen atom from LaNiO$_3$, yielding the reduced-RP LaNiO$_2$. The synthesis of RP nickelates has been studied extensively for decades \cite{Goodenough-1958,Wold1959176,DRENNAN1982621,327-1,327-2,Goodenough-1994,327-4,ZHANG1995236}. Among these, La$_2$NiO$_4$ and LaNiO$_3$ are the most stable and well-characterized phases, while La$_4$Ni$_3$O$_{10}$ and La$_3$Ni$_2$O$_7$ are more challenging to synthesize.

We want to organize this paper as follows: we begin with a brief overview of superconductivity in the hole-doped LaNiO$_2$ system (we will label it as the 112 system for short). We then focus on the bilayer La$_3$Ni$_2$O$_7$ (short for 327 system), highlighting both the high-pressure and thin-film superconducting phases. Following this, we examine the trilayer La$_4$Ni$_3$O$_{10}$ (short for 43(10) system) and other relative multilayer nickelates. Finally, we discuss current limitations in the field and outline future directions for research.

Before delving into detailed discussions, we would like to highlight an important point. Although multilayer nickelates exhibit superconductivity, they are fundamentally different from their cuprate counterparts. In multilayer cuprates, superconductivity primarily resides within individual CuO$_2$ planes, which are only weakly coupled. In contrast, the layers in multilayer nickelates are strongly coupled, indicating that superconductivity arises collectively in the multilayers.

\begin{figure*}
	\begin{center}
		\fig{7.0in}{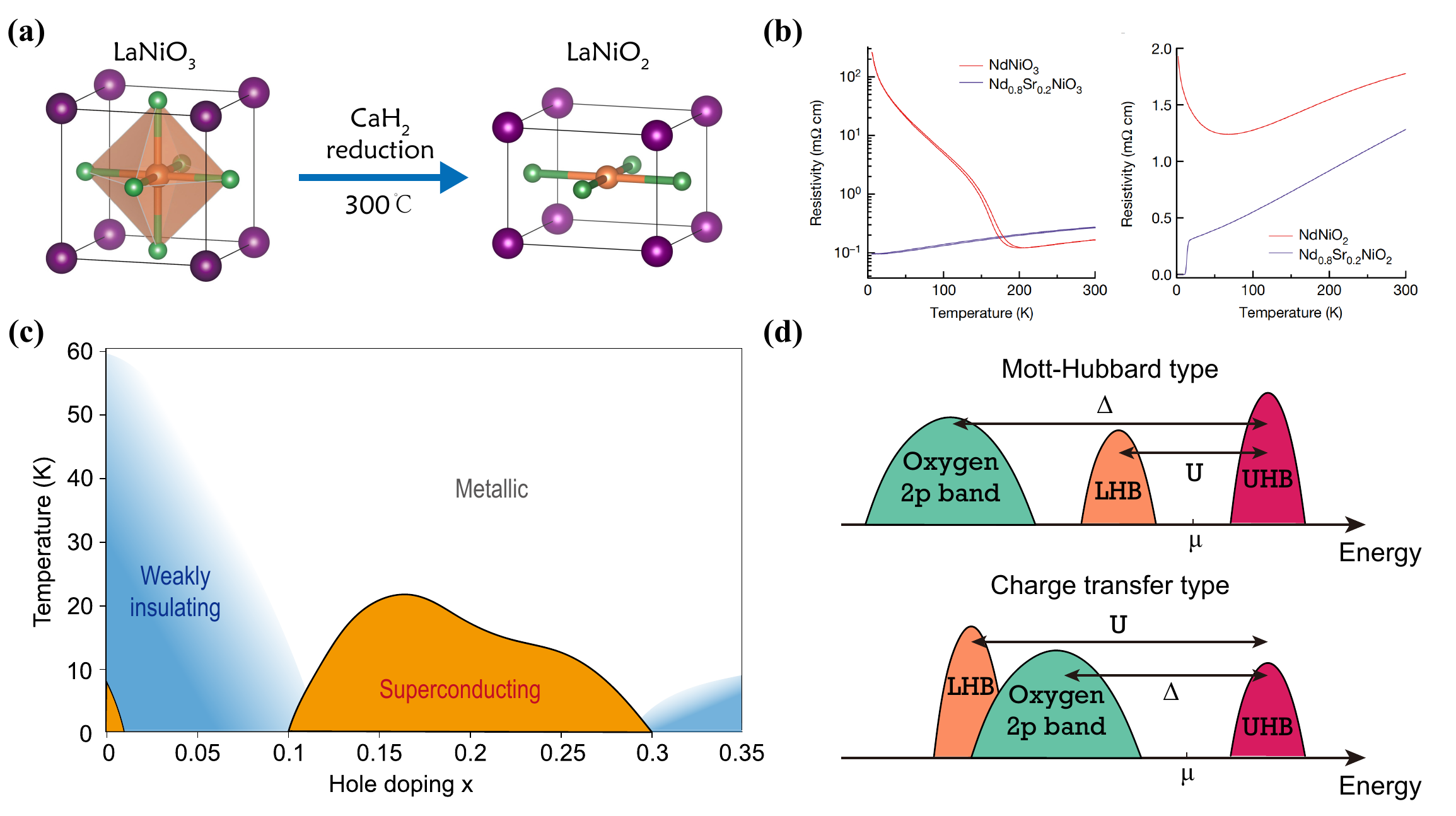}\caption{(a) LaNiO$_2$ is synthesized by reducing LaNiO$_3$ using CaH$_2$, which selectively removes oxygen atoms from the LaO planes \cite{lidanfeng}. (b) The left panel shows the resistivity of NdNiO$_3$ (insulating) and Nd$_{0.8}$Sr$_{0.2}$NiO$_3$ (metallic) thin films. The right panel displays the resistivity of the reduced compounds, NdNiO$_2$ and Nd$_{0.8}$Sr$_{0.2}$NiO$_2$. A superconducting transition with an onset at 14.9 K is observed in Nd$_{0.8}$Sr$_{0.2}$NiO$_2$ \cite{lidanfeng}. (c) The global phase diagram of hole-doped LaNiO$_2$ \cite{112-review-2}.
        (d) The “parent” state of LaNiO$_2$ is a Mott insulator, in contrast to cuprates, where the parent state is a charge-transfer insulator. We emphasize that the “parent” state here refers to the idealized insulating limit of LaNiO$_2$, which may differ from the actual, experimentally realized parent compound.
			\label{fig2}}
	\end{center}
	\vskip-0.5cm
\end{figure*}


\section{112}
Since several comprehensive reviews on LaNiO$_2$-based superconductors already exist \cite{112-review-1,112-review-2}, we provide only a brief overview here. As discussed earlier, Ni$^{1+}$ is both rare and thermodynamically unstable. The typical approach to stabilize it involves removing oxygen from the more stable LaNiO$_3$, as illustrated in Fig. \ref{fig2}(a). This reduction is commonly achieved via a soft-chemistry topotactic reaction using metal hydrides, a method applied to LaNiO$_2$ since 1999 \cite{sodium-hydride-1,sodium-hydride-2,cah2-1,cah2-2}. A major breakthrough came in 2019, when Li et al. successfully synthesized hole-doped Nd$_{0.8}$Sr$_{0.2}$NiO$_2$ superconducting thin films on SrTiO$_3$ substrates using CaH$_2$ as the reducing agent \cite{lidanfeng}. As shown in Fig. \ref{fig2}(b), the parent NdNiO$_3$ (a low-temperature-magnetic insulator) and the metallic Nd$_{0.8}$Sr$_{0.2}$NiO$_3$ films were first grown using pulsed-laser deposition. Subsequent reduction with CaH$_2$ yielded NdNiO$_2$, which exhibits a resistivity upturn below 70 K. In contrast, the hole-doped Nd$_{0.8}$Sr${_{0.2}}$NiO$_2$ enters a superconducting state below 14.9 K.

After six years of intensive research, a clearer picture of the hole-doped 112 phase diagram has emerged, as shown in Fig. \ref{fig2}(c). Similar to the cuprates, a superconducting dome appears in the doping range of approximately $x = 0.1$ to $x = 0.3$, with the superconducting transition temperature $T_c$ reaching up to 40 K \cite{secs_112,danfeng_112_secs}. This superconducting dome is flanked on both sides by weakly insulating phases, characterized by a low-temperature resistivity upturn—consistent with the behavior observed in undoped NdNiO$_2$, shown in Fig. \ref{fig2}(b). Interestingly, the $x = 0$ composition is not always insulating. Several studies have reported a new superconducting phase at $x = 0$ \cite{x0-PhysRevX.15.021048,x0-PrNiO2,x0-danfeng}, as highlighted in Fig. \ref{fig4}(a).

\begin{figure}
	\begin{center}
		\fig{3.4in}{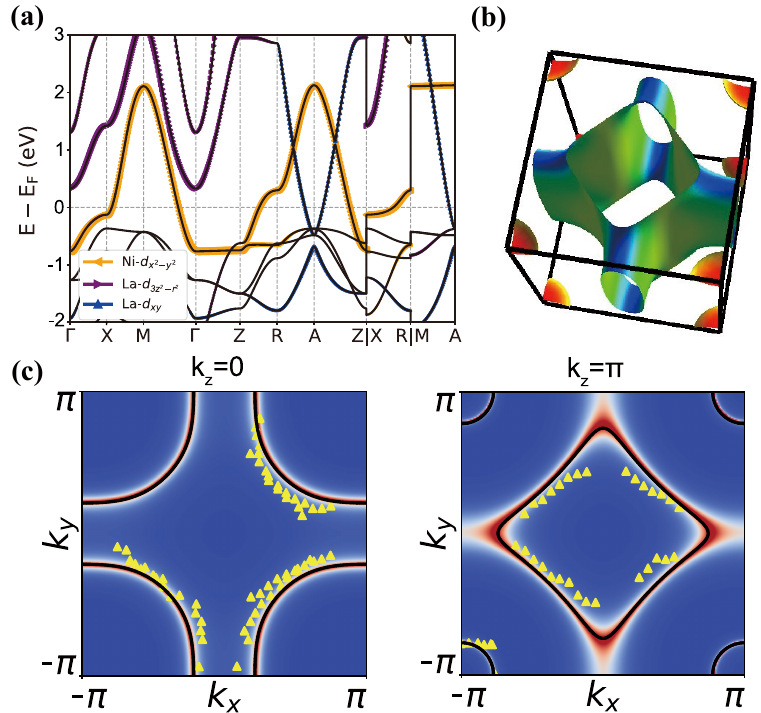}\caption{(a) The band structure of LaNiO$_2$, where the dominate valence electrons are from Ni 3$d_{x^2-y^2}$ and La 5$d$ orbitals.  (b) The three-dimensional Fermi surface contour of La$_{0.8}$Sr$_{0.2}$NiO$_2$. (c) The Fermi surfaces at $k_z = 0$ and $k_z = \pi$ calculated using tight-binding and dynamical mean-field theory \cite{chen2024electronicstructuresuperconductingproperties}. The yellow markers indicate ARPES experimental data points \cite{nieyf-sa}.
			\label{fig3}}
	\end{center}
	\vskip-0.5cm
\end{figure}

\begin{figure}
	\begin{center}
		\fig{3.4in}{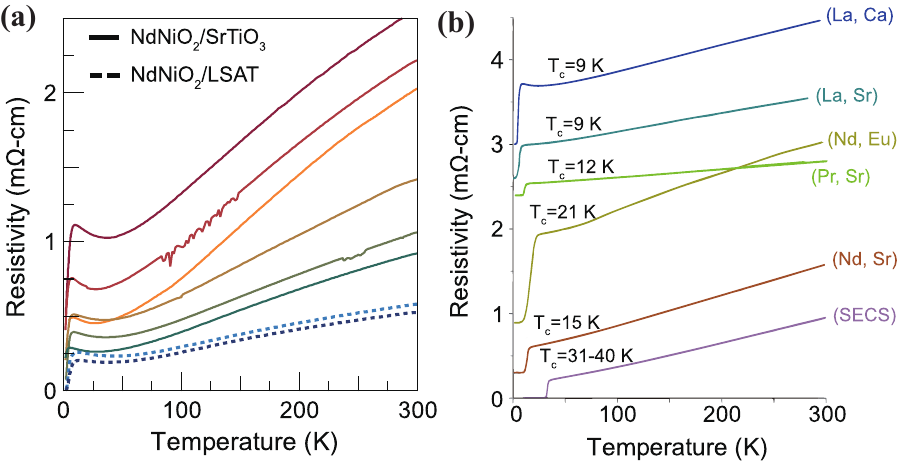}\caption{(a) Superconducting transition found in different NdNiO$_2$ samples \cite{x0-PhysRevX.15.021048}. (b) In comparison to other 112 SCs, Sm$_{1-x}$Sr$_x$NiO$_2$ co-doped with Eu and Ca (SECS) achieve higher transition temperatures, reaching up to $T_c \sim 40$ K \cite{secs_112}.
			\label{fig4}}
	\end{center}
	\vskip-0.5cm
\end{figure}

Although Ni$^{1+}$ shares the same $d^9$ electronic configuration as Cu$^{2+}$ in the cuprates, hole-doped LaNiO$_2$ differs from hole-doped cuprates in the following respects:
\begin{enumerate}
    \item The ideal insulating limit, or the “parent” state, LaNiO$_2$ is considered a Mott insulator rather than a charge-transfer insulator \cite{ZSA_PhysRevLett.55.418, Sawatzky_PhysRevLett.124.207004}. As illustrated in Fig. \ref{fig2}(d), in cuprate superconductors, the Cu $d$-orbitals split into a lower Hubbard band (LHB) and an upper Hubbard band (UHB), separated by the on-site Coulomb interaction energy $U$. The oxygen $p$ band lies between these two Hubbard bands, with a charge-transfer gap $\Delta_{CT}$ between the O $p$ levels and the UHB. As a result, doped holes in cuprates tend to reside on oxygen sites, forming the well-known Zhang-Rice singlet state \cite{zhang_rice_PhysRevB.37.3759}. In contrast, for LaNiO$_2$, the charge-transfer gap $\Delta_{CT}$ is significantly larger than the Hubbard $U$, which pushes the oxygen $p$ band below the LHB. Consequently, doped holes in LaNiO$_2$ preferentially occupy Ni $d$ orbitals rather than oxygen sites. This key distinction suggests that hole-doped LaNiO$_2$ may behave more like electron-doped cuprates rather than hole-doped ones.
    \item The electronic structure of LaNiO$_2$ exhibits pronounced $k_z$ dispersion. Thanks to advances in sample quality, angle-resolved photoemission spectroscopy (ARPES) measurements on hole-doped LaNiO$_2$ have become feasible \cite{nieyf-sa, donglai-nsr}. For instance, in La$_{0.8}$Sr$_{0.2}$NiO$_2$ (LSNO), the low-energy electronic states are primarily derived from Ni 3$d_{x^2-y^2}$ and La 5$d$ orbitals, as shown in Fig. \ref{fig3}(a). The Fermi surface (FS) contour at $k_z = 0$, plotted in Figs. \ref{fig3}(b-c), reveals a hole-like FS from the Ni 3$d_{x^2-y^2}$ orbital, closely resembling that of cuprates. However, this hole FS transforms into an electron-like FS at $k_z = \pi$, highlighting the strong three-dimensionality of the system. Since reduced dimensionality is believed to be a key factor contributing to the high $T_c$ in cuprates, the pronounced $k_z$ dispersion in LSNO may be one reason why its superconducting transition temperature remains lower in comparison \cite{chen2024electronicstructuresuperconductingproperties}.
    
    \item The role of 5$d$ electrons in LaNiO$_2$ remains a topic of ongoing debate. In addition to the Fermi surface derived from Ni 3$d$ electrons, ARPES measurements reveal a small Fermi pocket near the corner of the Brillouin zone (BZ), attributed to La 5$d$ electrons \cite{nieyf-sa, donglai-nsr}. Notably, the 5$d$ band shows no observable band renormalization, indicating that electronic correlations in this band are very weak. This weak correlation is consistent with expectations based on doping and band structure. Density functional theory (DFT) calculations show that the 5$d$ band has a large bandwidth of approximately 2 eV, and the observed Fermi pocket lies far from half-filling. As a result, the primary role of these weakly correlated 5$d$ electrons is to provide intrinsic hole doping to the Ni 3$d$ band. Consequently, LaNiO$_2$ is not a true parent compound in the sense of having a half-filled 3$d$ band, due to this self-doping effect from the 5$d$ states.
\end{enumerate}

From Fig. \ref{fig2}(c) phase diagram, there are two regions of weak insulator with a resistivity upturn. The origin of this upturn is widely debated. There are two major scenarios: Kondo scattering and the disorder effect. Theoretically, it was proposed that the 5$d$ electron contributes to conductivity, which is Kondo-scattered by the localized Ni 3$d$ electrons at low temperature \cite{kondo_PhysRevB.101.020501}. On the other hand, one should notice that topotactic reduction always introduces various defects to LaNiO$_2$. This kind of disorder effect is unavoidable for the 112 system. It has been shown that by improving sample quality, the upturn behavior shows a clear dependence on the disorder level in the overdoped region \cite{Hwang-WI}.

Finally, we highlight two recent advances in the 112 nickelate systems:
\begin{itemize}
    \item[(a)] Superconductivity at $x = 0$: In a study conducted a few years ago, signs of superconductivity were already observed in undoped LaNiO$_2$ \cite{osada2021nickelate}. Recently, superconductivity with $T_c \approx 11$ K has been definitively observed in the nominally undoped “parent” compounds NdNiO$_2$ and PrNiO$_2$ \cite{x0-PhysRevX.15.021048, x0-PrNiO2, x0-danfeng}, adding a new feature to the 112 phase diagram. As shown in Fig. \ref{fig4}(a), improvements in substrate selection and sample quality for NdNiO$_2$ have led to significantly reduced residual resistivity. While some samples still display resistivity upturns, superconducting transitions are clearly observed. This finding underscores the crucial role of disorder in determining the physical properties of 112 systems. It also raises important questions: Is the $x=0$ superconductivity intrinsic? How is it connected to the main superconducting dome? 
    \item[(b)] Enhancement of $T_c$ via chemical tuning: Raising $T_c$ remains a key goal in 112 research. Carrier doping and epitaxial strain from substrates are known to significantly influence superconductivity in thin films. Recently, Sm$_{1-x}$Sr$_x$NiO$_2$ and co-doped variants incorporating Eu and Ca have achieved transition temperatures approaching 40 K \cite{secs_112,danfeng_112_secs}, as shown in Fig. \ref{fig4}(b). These co-doped compounds exhibit remarkably low resistivity—on the order of 0.01 m $\Omega \cdot$cm—and a record-small $c$-axis lattice parameter. The origin of this improvement remains unclear, calling for further experimental and theoretical investigation.
\end{itemize}


\begin{figure*}
	\begin{center}
		\fig{7.0in}{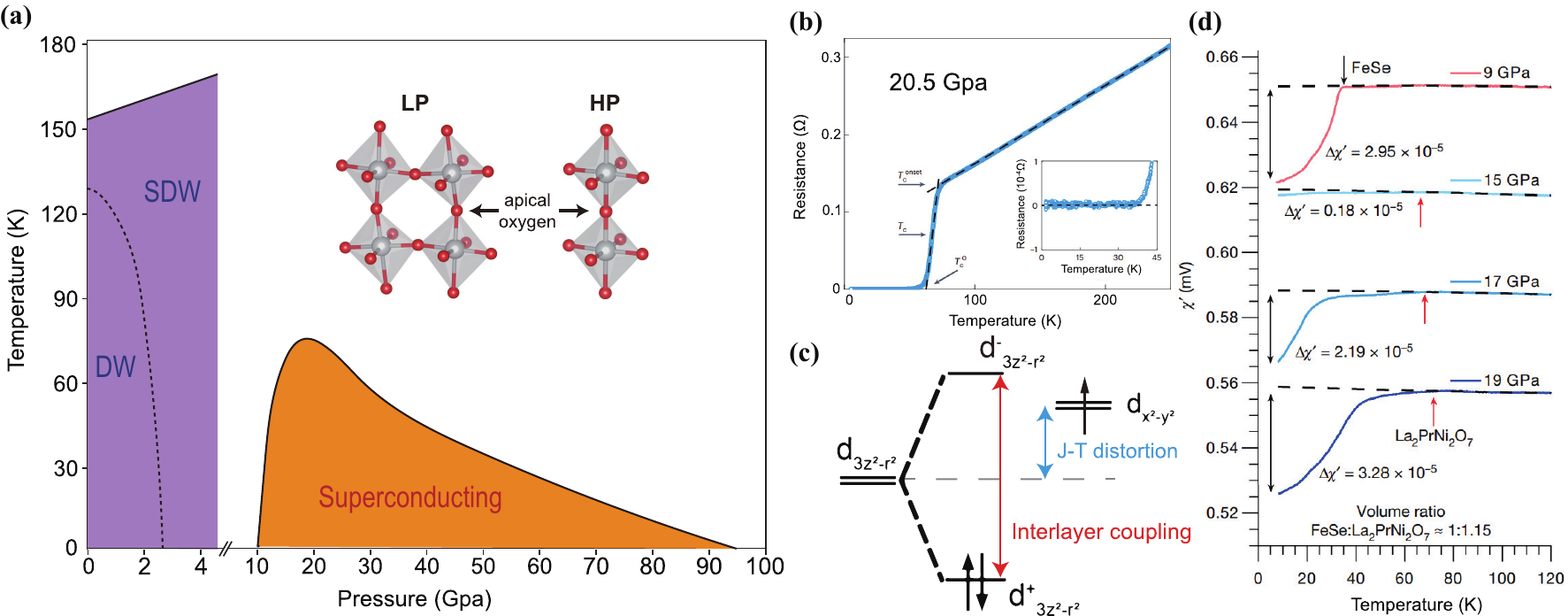}\caption{
        (a) Global phase diagram of La$_3$Ni$_2$O$_7$ as a function of pressure and temperature, revealing a transition from a low-pressure (LP) phase to a high-pressure (HP) phase. The inset illustrates the distinct crystal structures associated with each phase. The apical oxygen, which connects the two NiO$_2$ layers, plays a key role in the electronic properties of the 327 compound.
        (b) Temperature-dependent resistance $R(T)$ of La$_3$Ni$_2$O$_7$ at 20.5 GPa, showing a superconducting transition onset at 66 K and reaching zero resistance around 40 K \cite{yuanhq_np}.
        (c) Schematic of the $e_g$ orbital energy levels in bilayer La$_3$Ni$_2$O$_7$. Due to Jahn-Teller (J-T) distortion, the $d_{3z^2-r^2}$ orbital lies below the $d_{x^2-y^2}$ orbital. Strong interlayer coupling further splits the $d_{3z^2-r^2}$ states into bonding ($d_{3z^2-r^2}^+$) and antibonding ($d_{3z^2-r^2}^-$) orbitals.
        (d) The a.c. magnetic susceptibility $\chi'(T)$ of La$_2$PrNi$_3$O$_7$ under various pressures of up to 19 GPa. The dashed line represents the background extrapolated from the high-temperature region \cite{chengjg_nature}. The superconducting shielding volume fraction can reach more than $90\%$.
			\label{fig5}}
	\end{center}
	\vskip-0.5cm
\end{figure*}

In summary, the 112 nickelates represent a new class of superconductors based on the 3$d^9$ configuration, analogous to hole-doped cuprates. While 5$d$ electrons introduce additional complexity, the low-energy physics is still primarily governed by the Ni 3$d_{x^2 - y^2}$ orbitals. Despite some notable differences, many key concepts from cuprate superconductivity are transferable to the 112 systems. In particular, the widely anticipated nodal $d$-wave pairing symmetry remains a central prediction awaiting experimental confirmations \cite{112-d-1,112-d-2,112-d-3,112-d-4,112-d-5}.

\section{327}

Following the discovery of superconductivity in hole-doped 3$d^9$ 112 nickelates, pressurized La$_3$Ni$_2$O$_7$ was identified in 2023 as a superconductor with an onset $T_c$ near 80 K \cite{meng_wang}. 
The compound 327 has been successfully synthesized for several decades \cite{327-1,327-2,Goodenough-1994,327-4}.
As highlighted in Fig. \ref{fig1}, La$_3$Ni$_2$O$_7$ possesses a distinct valence electron configuration of 3$d^{7.5}$, differing from the 3$d^{9}$ configuration of the 112 nickelates. This establishes La$_3$Ni$_2$O$_7$ as a novel class of nickelate superconductor. While a previous review details the early development of La$_3$Ni$_2$O$_7$ properties \cite{327-review}, this work aims to describe the system from a different perspective. Readers seeking information on its history, synthesis, and other aspects are directed to Ref. \cite{327-review}. 
We also want to note that owing to the crystal quality of La$_3$Ni$_2$O$_7$ or polycrystalline sample forms, different references may arrive at different conclusions. 

\subsection{Phase diagram and high-pressure superconductivity}

The global phase diagram as function of pressure and temperature is plotted in Fig. \ref{fig5} (a). 
The phase diagram contains two parts: the low-pressure (LP) region and the high-pressure (HP) region. The major difference between the LP phase and the HP phase comes from their structure. As described in Fig. \ref{fig1}, the central ingredient of La$_3$Ni$_2$O$_7$ structure is the bilayer NiO$_2$ planes. Each bilayer is formed by two shared apical oxygen NiO$_6$ octahedra. This shared apical oxygen has great importance in the physical properties of La$_3$Ni$_2$O$_7$.
In the HP phase, two NiO$_6$ octahedra are lined up, making the bond angle between top Ni-apical O and bottom Ni-apical O exactly 180$^{\circ}$. The space group of HP phase is \textit{Fmmm} \cite{meng_wang} or \textit{I4/mmm} under higher pressure \cite{li2025identification} with 2 Ni per unit cell (1 in each layer). On the other hand, the bond angle tilts away to 168$^{\circ}$ in the LP phase. And the nearest-neighbor octahedra tilting angle alternates as the $(\pi,\pi)$ Neel order. The space group of LP becomes \textit{Amam} with 4 Ni per unit cell (2 Ni in each layer). 


In the HP phase, a superconducting dome emerges around 10 GPa, reaches its maximum $T_c$ near 20 GPa, and gradually disappears around 90 GPa. As illustrated in Fig. \ref{fig5}(b), the electrical resistance $R(T)$ measured at 20.5 GPa using a diamond anvil cell (DAC) shows a clear superconducting transition beginning at 66 K ($T_c$), with zero resistance reached at approximately 40 K ($T_c^0$) \cite{yuanhq_np}. A definitive signature of superconductivity is the Meissner effect, which confirms bulk superconductivity through the observation of diamagnetism. However, due to limited sample quality, measurements of the magnetic response and superconducting volume fraction in the HP phase have been subject to significant debate \cite{sunll-meissner, Hepting_PhysRevLett.133.146002}.
This debate has been substantially addressed by recent experiments on La$_2$PrNi$_2$O$_7$ \cite{chengjg_nature}. As shown in Fig. \ref{fig5}(d), the a.c. magnetic susceptibility $\chi'(T)$ was measured using the mutual induction method in a multianvil press, with a FeSe single crystal used as a calibration reference. At 9 GPa, FeSe shows a clear superconducting diamagnetic signal below 30 K. Notably, FeSe is known to transition into a non-superconducting hexagonal phase above 10 GPa, eliminating its diamagnetic response. In contrast, La$_2$PrNi$_2$O$_7$ begins to exhibit strong diamagnetic signals above 15 GPa. By comparing $\chi'(T)$ between the two samples, the superconducting shielding volume fraction in La$_2$PrNi$_2$O$_7$ was estimated to exceed 90\%, providing compelling evidence for bulk superconductivity in the HP phase.


Beyond the HP superconducting phase, the LP phase of La$_3$Ni$_2$O$_7$ also exhibits intriguing behavior. 
The LP phase undergoes a density wave transition around 110 K and 153 K from resistance kinks \cite{meng-wang-scpma,yuanhq_np}.
This transition appears to be weak and is nearly undetectable in heat capacity measurements \cite{meng-wang-scpma}. With increasing pressure, the temperature of this resistance kink systematically decreases. In parallel, various spectroscopic and scattering techniques, including nuclear magnetic resonance (NMR), resonant inelastic X-ray scattering (RIXS), muon spin relaxation ($\mu$SR), and neutron diffraction, have reported signatures consistent with the onset of a spin density wave (SDW) state near 150 K \cite{musr, rixs, nmr, plokhikh_unraveling_2025}. However, the precise magnetic structure of La$_3$Ni$_2$O$_7$ remains under debate. Interestingly, the SDW transition temperature ($T_{\mathrm{SDW}}$) increases with applied pressure, while the resistance kink decreases, suggesting the presence of two distinct density wave transitions. It is widely hypothesized that, in addition to the SDW, a charge density wave (CDW) transition also occurs. Clarifying the interplay between these two density wave orders is essential for understanding the complex phase diagram of La$_3$Ni$_2$O$_7$.


The transition from the LP to HP crystal structure in La$_3$Ni$_2$O$_7$ has been identified by synchrotron \textit{X}-ray diffraction to occur between 9 and 11 GPa \cite{meng_wang,chengjg_nature}, and is likely of first-order nature. Notably, the pressure required to induce superconductivity in this system is relatively modest, especially when compared to extremely high-pressure superconductors such as H$_3$S. In fact, superconductivity in polycrystalline La$_3$Ni$_2$O$_7$ samples has been observed at pressures as low as 6 GPa \cite{chengjg_PhysRevX.14.011040}. These observations suggest that conventional electron-phonon coupling is unlikely to be the primary mechanism behind the high superconducting transition temperature. Instead, the key factor appears to be the stabilization of the HP crystal structure itself. This idea is further supported by recent progress in La$_3$Ni$_2$O$_7$ thin films, where similar structural stabilization strategies have been employed to approach the superconducting phase \cite{hwang}.

\subsection{Electronic structure and apical oxygen vacancy}
In this subsection, we discuss the electronic structure of La$_3$Ni$_2$O$_7$ in its HP structure. The LP phase can be understood from band folding owing to its unit cell doubling. In analogy to the single CuO$_2$ plane in cuprates, the electronic structure of the bilayer NiO$_2$ plane plays the key role. From the quantum chemistry perspective, two Ni$^{2.5+}$ atoms, under an octahedral crystal field, have fully filled $t_{2g}$ orbitals, and additional 3 electrons filling the $e_{g}$ orbitals, as illustrated in Fig. \ref{fig5}(c). The apical oxygen mediated interlayer hopping significantly split the $d_{3z^2-r^2}$ orbtial into bonding ($d_{3z^2-r^2}^+$) and antibonding ($d_{3z^2-r^2}^-$) orbitals. $d_{x^2-y^2}$ interlayer hopping is relatively weaker. 
On the other hand, the Jahn–Teller distortion also contributes to raising the energy of the $d_{x^2-y^2}$ orbital relative to the $d_{3z^2 - r^2}$ orbital. Hence, it tends to fully occupy the $d_{3z^2-r^2}^+$ band and quarter fill the two $d_{x^2-y^2}$ orbitals based on the atomic energy levels.

\begin{figure*}
	\begin{center}
		\fig{7.0in}{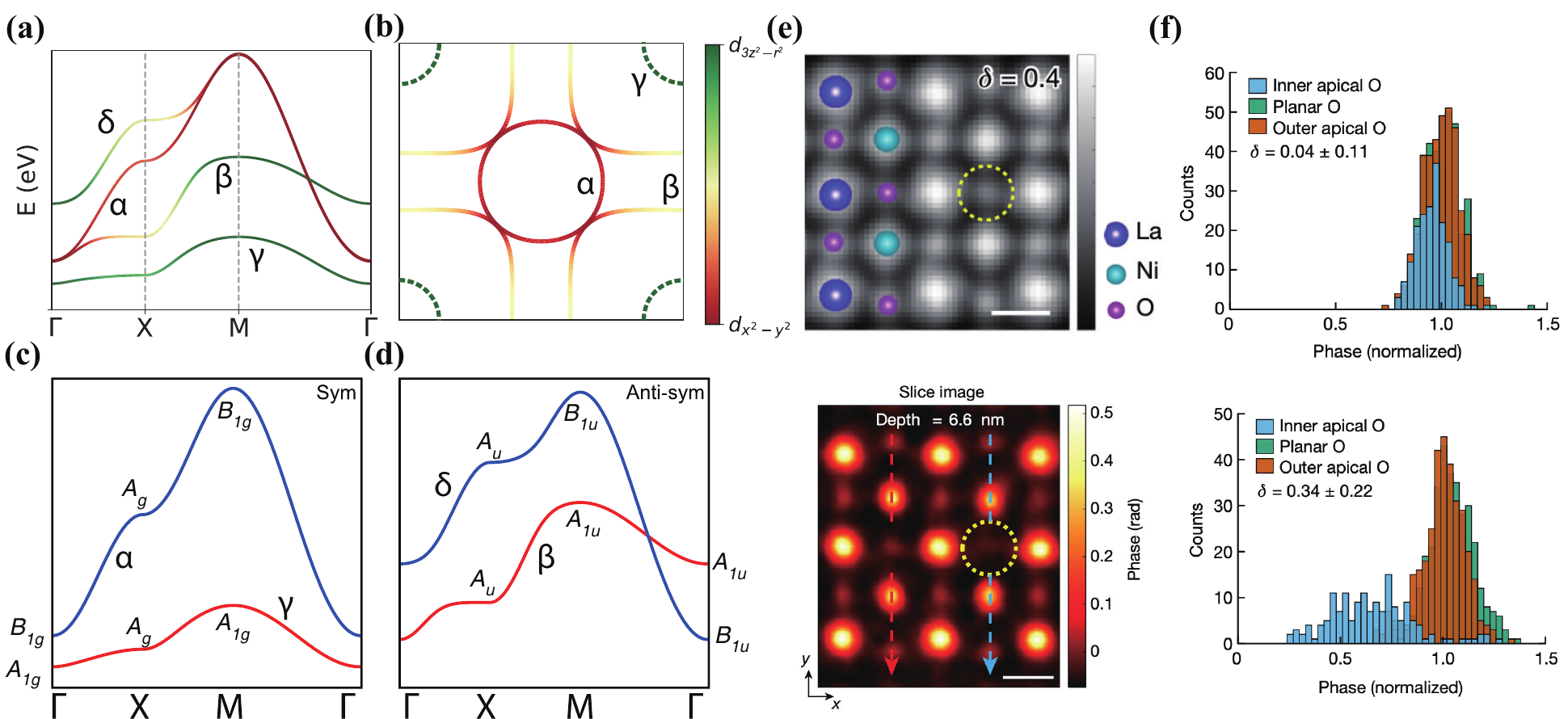}\caption{(a) The HP band structure of La$_3$Ni$_2$O$_7$ and its projections into $e_g$ orbitals. %
        (b) Fermi surfaces of La$_3$Ni$_2$O$_7$, featuring two prominent sheets: the $\beta$-FS and the $\alpha$-FS. The presence of a $\gamma$-FS at the BZ corner crossing the Fermi level remains under debate \cite{wangyx_PhysRevB.110.205122, Wangyx_2025}. (c) The four low-energy bands of La$_3$Ni$_2$O$_7$ can be categorized into symmetric (Sym) and antisymmetric (Anti-Sym) sectors. Their irreducible representations have been systematically analyzed \cite{Wangyx_2025}. Note that the two bands in the Anti-Sym sector are strongly entangled. (e) Simulated and experimentally observed STEM structures of La$_3$Ni$_2$O$_7$ \cite{chenzhen-TEM}. Yellow circles indicate inner apical oxygen vacancies. (f) Phase histograms corresponding to different oxygen sites in regions with $\delta = 0.04$ and $\delta = 0.34$, normalized to the average phase of the outer apical oxygen sites \cite{chenzhen-TEM}. A significant loss of inner apical oxygens is observed in the $\delta = 0.34$ region.
			\label{fig6}}
	\end{center}
	\vskip-0.5cm
\end{figure*}

From the band structure perspective and density functional theory (DFT) calculation, the low-energy physics is formed by the bilayer $e_{g}$ orbitals. Therefore, one can use the bilayer $d_{x^2-y^2}$ and $d_{3z^2-r^2}$ to construct a two-orbital bilayer tight-binding (TB) model \cite{yaodx}. Their band structure and orbital contents are plotted in Fig.\ref{fig6}(a). Generally speaking, there are four sets of bands, which are normally named as $\alpha$, $\beta$, $\gamma$ and $\delta$. However, we want to remind readers that DFT calculations in strongly correlated systems highly depend on the way treating correlation and exchange-correlation functionals \cite{wangyx_PhysRevB.110.205122,Wangyx_2025}. Although the topography of the band structure is similar, the Fermi surface (FS) features vary on the different methods as illustrated in Fig.\ref{fig6}(b). For example, the $\gamma$ band becomes fully occupied using hybrid functionals instead of a hole pocket using generalized gradient approximation (GGA) functionals \cite{wangyx_PhysRevB.110.205122,Wangyx_2025}. The only way to justify these calculations is the experimental findings. Especially, FS is always the key information for any superconductivity. It is originally thought that the high density of states $\gamma$ pocket is the key factor for realizing superconductivity in La$_3$Ni$_2$O$_7$ \cite{meng_wang}. But decisive experiments under high pressure are needed to support this conjecture originated from DFT calculations.

From the symmetry point of view, the four bands can be further split into two sectors: inversion symmetric ($\alpha$ and $\gamma$) and inversion antisymmetric ($\beta$ and $\delta$). One can do irreducible representation analyses for each sector.
For the inversion symmetric case in Fig. \ref{fig6}(c), we can find that the $\alpha$ band and $\gamma$ are well separated. Therefore, each band in the inversion symmetric sector can be Wannierized separately. The Wannierization is more complicated in inversion antisymmetric sector \cite{Wangyx_2025}. As plotted in Fig. \ref{fig6}(d), although the $A_{1u}$ band is slightly above the $B_{1u}$ band at $\Gamma$, they switch positions at M point. Hence, the inversion antisymmetric bands are highly entangled and one cannot write down an isolated Wannier function for each band separately. This feature indicates that although $\beta$-FS mostly contains $d_{x^2-y^2}$, the $d_{3z^2-r^2}$ component can not be ignored from a Wannierization perspective \cite{Wangyx_2025}.

From the above discussion, we can find that the apical oxygen plays an important role in the electronic structure of La$_3$Ni$_2$O$_7$. Especially, the coupling between two Ni layers is through this apical oxygen.
Defects, especially oxygen vacancies, are unavoidable factors in oxide sample growth. 
Experimentally, the scanning transmission electron microscopy (STEM) has been applied to study defect structures in La$_3$Ni$_2$O$_{7-\delta}$ \cite{chenzhen-TEM}.
STEM can directly image the electrostatic potential of atoms through the phase information. As plotted in Fig. \ref{fig6}(e), phase image of a slice can probe the absence of inner apical oxygens with the yellow dashed circles.  From the statistics of oxygen contents at different positions in Fig.\ref{fig6}(f), the inner apical oxygens become much less in $\delta=0.34$ region than the $\delta=0$ region from the same La$_3$Ni$_2$O$_{7-\delta}$ sample. Hence, the inner apical oxygen vacancies are major defects and the La$_3$Ni$_2$O$_{7-\delta}$ sample shows a large inhomogeneity. Theoretically, there have already few attempts to address this problem. We hope for more findings to clarify this vacancy influence. 

\begin{figure*}
	\begin{center}
		\fig{7.0in}{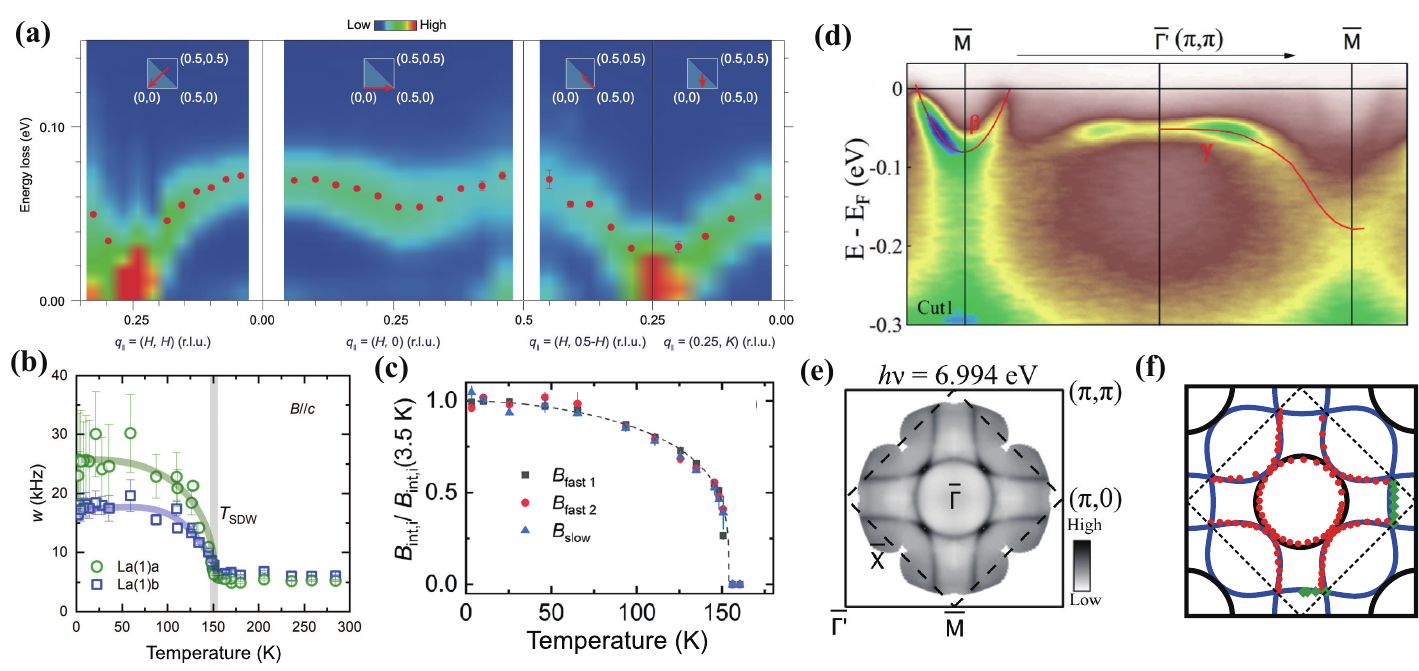}\caption{(a) RIXS intensity maps along high-symmetry directions in low-pressure (LP) La$_3$Ni$_2$O$_7$, with red-filled circles marking the peak positions of magnetic excitations. Notably, the magnetic excitation spectrum becomes gapless at $Q = (\pi/2, \pi/2)$ \cite{rixs}. (b) Temperature-dependent NMR linewidths for the La(1)a and La(1)b sites, indicating a SDW transition at 150 K in LP La$_3$Ni$_2$O$_7$ \cite{nmr}. (c) Temperature evolution of the magnetic order parameter measured via $\mu$SR, confirming magnetic ordering in the LP phase \cite{musr}. (d) ARPES measurements of the band structure in LP La$_3$Ni$_2$O$_7$ \cite{zhouxj}. (e) Laser-based ARPES mapping of the Fermi surface in the same LP phase \cite{zhouxj}. (f) Fermi surface calculated using hybrid functional DFT, showing excellent agreement with the experimental ARPES data in (e), as indicated by the red and green points \cite{wangyx_PhysRevB.110.205122}.
			\label{fig7}}
	\end{center}
	\vskip-0.5cm
\end{figure*}

\subsection{Density waves in 327}
Although experimental measurements on the 327 phase under high pressure remain challenging, conventional material characterization techniques can be extensively applied to the LP phase. Methods such as NMR, RIXS, and ARPES have been instrumental in probing the physical properties of the 327 system.

Initial $^{139}$La NMR measurements on polycrystalline La$_3$Ni$_2$O$_{7}$ samples suggested a possible density-wave-like transition below 150 K \cite{kakoi_multiband_2024,fukamachi_139nmr_2001,fukamachi_studies_2001}. However, whether this transition corresponds to a SDW or a CDW remained unclear. Recent studies employing $\mu$SR, NMR, RIXS measurements, and neutron scattering, have identified an SDW transition in  La$_3$Ni$_2$O$_{7}$  with a transition temperature $T_{\mathrm{SDW}}$ around 150 K. Zero-field $\mu$SR experiments on polycrystalline samples have confirmed magnetic order below 154 K \cite{musr}. Furthermore, RIXS measurements revealed strongly dispersive magnetic excitations, which exhibit softening towards zero energy at the wavevector (0.25, 0.25) \cite{rixs}, as shown in Fig. \ref{fig7}(a). This soft-mode behavior is a hallmark of translation symmetry breaking and is consistent with an SDW order along the ($\pi$,$\pi$) direction. 
In parallel, the temperature-dependent $^{139}$La NMR spectra and nuclear spin-lattice relaxation rate ($1/T_1$) in single crystal provide clear evidence for SDW ordering below around 150 K \cite{nmr}, as shown in Fig. \ref{fig7}(b). Intriguingly, when pressure is applied, both NMR and $\mu$SR measurements show that $T_{SDW}$ increases, a trend opposite to that observed in transport measurements where pressure suppresses this phase \cite{nmr,musr2}. This apparent discrepancy implies that the transition observed in transport may originate from a different kind of density-wave order, distinct from the SDW state. In addition to the SDW, the existence of a CDW order has been proposed. Recent optical conductivity measurements revealed opening of an energy gap below 115 K in single crystal \cite{liu2024electronic}, which may suggest the formation of a CDW order. 

Although the SDW transition around 150 K has been confirmed, its exact magnetic structure is not yet fully determined and under debate. Nevertheless, a consensus from various experimental evidence points towards a “stripe-like” magnetic order. Specifically, RIXS studies have proposed two possibilities: a spin-charge stripe order or a double spin-stripe order. $\mu$SR data from polycrystalline samples are qualitatively consistent with the spin-charge stripe scenario \cite{musr}.  In single crystal, the anisotropic splitting in the $^{139}$La NMR spectra suggests the formation of a possible double spin-stripe with magnetic moments aligned along the $c$-axis \cite{nmr}. Interpretation of nuclear quadrupole resonance (NQR) data on polycrystalline samples remain divergent. Yashima et al. suggests a single SDW transition corresponding to a spin-charge stripe order \cite{yashima_microscopic_2025}, while Luo et al. reports the simultaneous occurrence of both SDW and CDW transitions near 150 K \cite{luo_microscopic_2025}. More recently, neutron scattering on polycrystalline samples indicates the existence of two distinct spin-charge stripe orders. This model features alternating large and small magnetic moments within the Ni-O layer, which are then stacked antiferromagnetically along the c-axis, characterized by two vectors: $q_1 = (\pi/2,\pi/2,0)$ and $q_2 = (\pi/2,\pi/2,\pi)$, corresponding to coexisting two different magnetic configurations in the system \cite{plokhikh_unraveling_2025}. A major reason for the current uncertainty surrounding the magnetic structure may lie in the inhomogeneity of currently available samples. $^{139}$La NMR and NQR studies have revealed that there are two distinct chemical sites for lanthanum atoms, both for those located between the Ni-O bilayers and for those outside the bilayers \cite{nmr}. This strongly implies the coexistence of two different structural phases within the systems, which introduces an additional complexity to the experimental determination of its intrinsic magnetic structure. Therefore, further investigations using more homogeneous single crystals are imperative for accurately determining the magnetic configuration.

Both synchrotron-based and laser-based ARPES have been applied to LP phase \cite{zhouxj}, as plotted in Fig. \ref{fig7}(d).
The band structure of LP-327 has similar components to the HP phase. But owing to the octahedra tilting, the translation symmetry is broken to $\sqrt{2}\times\sqrt {2}$ of HP. Any eigenenergy $E(k+Q)$ with $Q=(\pi,\pi)$ in the HP BZ is folded to $E(k)$ in the new smaller BZ. Crossing points of the original bands and folded bands open gaps. From Fig. \ref{fig7}(d), we can see the $\gamma$ band lies below the Fermi level with band top around -$50$ meV. Both $\beta$ and $\alpha$ bands are observed with renormalization. 
For FS, the $\beta$ FS center around M is folded to centering around the $\Gamma$ point, resulting in a topography in Fig. \ref{fig7}(e). The FS feature and electronic structure in HP can be well captured by a hybrid functional DFT calculation and its further correlation method Fig. \ref{fig7}(f) \cite{wangyx_PhysRevB.110.205122}.

\subsection{327 thin film}

\begin{figure} 
	\begin{center}
		\fig{3.6in}{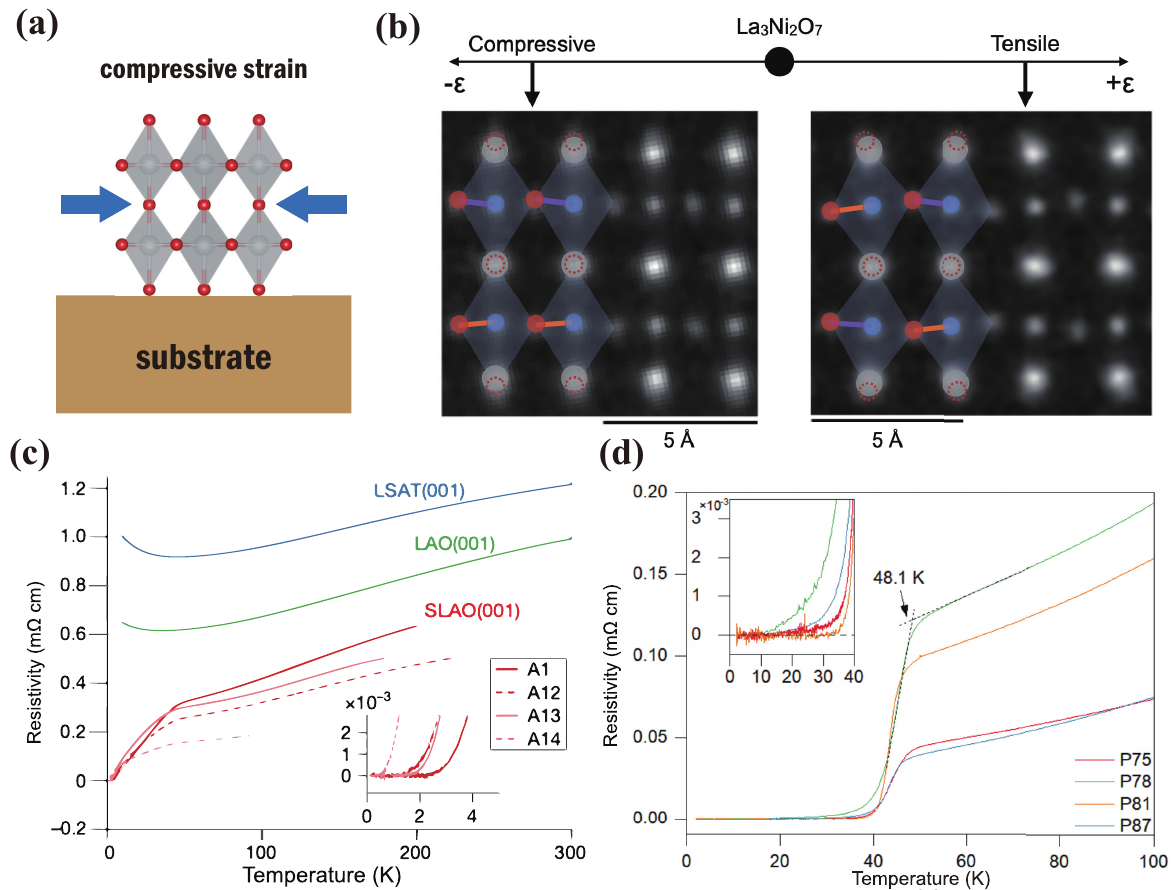}\caption{(a) Schematic illustration of the compressive strain induced by the substrate during thin-film deposition. (b) STEM images of La$_3$Ni$_2$O$_7$ thin films grown under compressive and tensile strain conditions \cite{327-film-structure}. (c) Temperature-dependent resistivity of La$_3$Ni$_2$O$_7$ thin films on various substrates \cite{hwang}. Films grown on compressive SLAO substrates exhibit a broad superconducting transition. A1, A12, A13, and A14 denote different sample identifiers. (d) Resistivity measurements of La$_2$PrNi$_2$O$_7$ thin films, showing a superconducting transition onset above 48.1 K. Sample labels include P75, P78, P81, and P87 \cite{la2prni2o7-film}.
			\label{fig8}}
	\end{center}
	\vskip-0.5cm
\end{figure}

Following the discovery of high-$T_c$ SC in La$_3$Ni$_2$O$_7$ under high pressure, numerous efforts have been made to synthesize superconducting thin films at ambient pressure. As previously discussed, stabilizing the La$_3$Ni$_2$O$_7$ structure is essential for the emergence of superconductivity. It is well established that compressive or tensile strain effectively mimics the key electronic structure modifications induced by high pressure \cite{erjia}. As such, strain offers an alternative pathway to stabilize the high-pressure phase of the 327 compound, as illustrated in Fig. \ref{fig8}(a). After extensive work using SrLaAlO$_4$ (SLAO) substrates, several groups have successfully reported superconducting 327-based films: Eun-Kyo Ko et al. reported superconducting 327 films \cite{hwang}; Guangdi Zhou et al. demonstrated superconductivity in La$_{2.85}$Pr$_{0.15}$Ni$_2$O$_7$ \cite{chen_zhuoyu}; Yidi Liu et al. reported superconducting La$_2$PrNi$_2$O$_7$ films \cite{la2prni2o7-film}; Bo Hao et al. observed superconductivity in La$_{3-x}$Sr$_x$Ni$_2$O$_7$ films \cite{yuefeng-sr327}; and Baiyang Wang et al. successfully grew La$_2$PrNi$_2$O$_7$ films using the oxide MBE method \cite{zxshen-arpes}, a resistivity drop was also observed in the transport measurements.

The temperature-dependent resistivity $\rho(T)$ of La$_3$Ni$_2$O$_7$ thin films grown on different substrates is shown in Fig. \ref{fig8}(c). On SLAO substrates, $\rho(T)$ exhibits a drop below 42 K, reaching zero resistance at approximately 2 K ($T_c^0$). In contrast, films grown on LaAlO$_3$ (LAO) and (LaAlO$_3$)$_{0.3}$(Sr$_2$TaAlO$_6$)$_{0.7}$ (LSAT) show a resistivity upturn below roughly 40 K, without a zero-resistance transition. Structurally, the strain imposed on the film by the substrate can be quantified as $\epsilon = \frac{a_s - a_f}{a_f}$, where $a_s$ and $a_f$ are the in-plane lattice constants of the substrate and the film without substrate constraints, respectively. The bulk in-plane lattice parameters can be used as an approximation for $a_f$. The in-plane pseudotetragonal lattice constants ($a$ and $b$ axes) of bulk 327 are approximately 3.83 \AA. SLAO, with $a_s = 3.75$ \AA, introduces about $-2\%$ in-plane compressive strain \cite{327-film-structure}. Multislice electron ptychography has been used to study the atomic-scale structural evolution of 327 films under varying strain conditions induced by different substrates, as shown in Fig. \ref{fig8}(b) \cite{327-film-structure}. These results demonstrate that compressive strain ($\epsilon < 0$) favors the stabilization of the HP phase, while tensile strain ($\epsilon > 0$) tends to preserve the LP structure, as expected.

\begin{figure}
	\begin{center}
		\fig{3.5in}{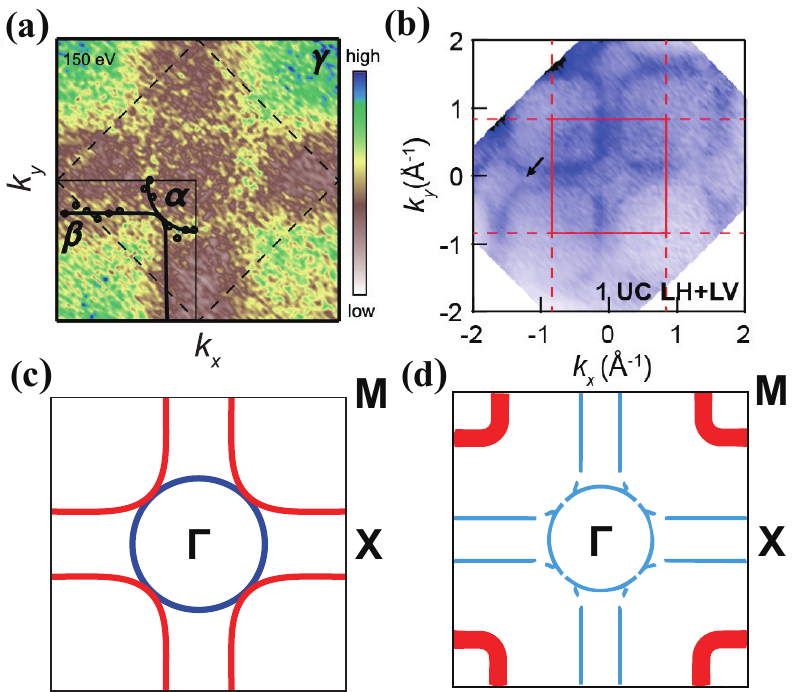}\caption{(a, c) ARPES-measured and schematic Fermi surfaces of La$_2$PrNi$_2$O$_7$ \cite{zxshen-arpes}, showing that the $\gamma$ band does not cross the Fermi level. (b, d) ARPES-measured and schematic Fermi surfaces of La$_{2.85}$Pr$_{0.15}$Ni$_2$O$_7$ \cite{junfeng-arpes1}, where the $\gamma$ band crosses the Fermi level.
			\label{fig9}}
	\end{center}
	\vskip-0.5cm
\end{figure}

Compared to the bulk structure under pressure, the $c$-axis of the 327 thin film expands to 20.6 \AA  \cite{327-film-structure}. Our DFT calculations reveal that the Jahn–Teller distortion is more sensitive to structural changes than interlayer coupling. As shown in Fig.~\ref{fig5}(c), the Jahn–Teller distortion, like interlayer coupling, lowers the energy of the bonding $d_{3z^2 - r^2}^+$ orbital. Therefore, both effects should be treated on equal footing.



Another critical factor in realizing superconductivity in 327 thin films is ozone annealing. It has been widely reported that many as-grown 327 films are insulating prior to annealing, indicating that disorder, inhomogeneity, and structural defects are unavoidable considerations in the superconducting behavior of these films. In particular, apical oxygen vacancies are believed to play an important role in shaping the electronic properties. Recent studies suggest that planar oxygen vacancies become predominant in superconducting 327 samples \cite{yuefeng-sr327}.

The development of superconducting 327 films has progressed rapidly. Notably, substantial improvements have been achieved in La$_2$PrNi$_2$O$_7$ films \cite{la2prni2o7-film}. As shown in Fig. \ref{fig8}(d), the maximum onset temperature $T_c$ reaches 48.1 K, with zero resistance achieved at temperatures exceeding 30 K. Interestingly, the normal-state transport in La$_2$PrNi$_2$O$_7$ exhibits clear Fermi-liquid behavior, and its Hall resistivity closely resembles that of overdoped La$_{1.75}$Sr$_{0.25}$CuO$_4$ \cite{la2prni2o7-film,327-fermi-liquid}. More recently, a superconducting dome has also been observed in Sr-doped La$_{3-x}$Sr$_x$Ni$_2$O$_7$ thin films, further establishing the tunability and richness of the 327 film phase diagram \cite{yuefeng-sr327}. Additionally, applying high pressure to strain-engineered 327 thin films has further enhanced transition temperature, reaching a maximum T$_c$ of approximately 60 K \cite{327-film-pressure}.


\begin{figure*}
	\begin{center}
		\fig{7.0in}{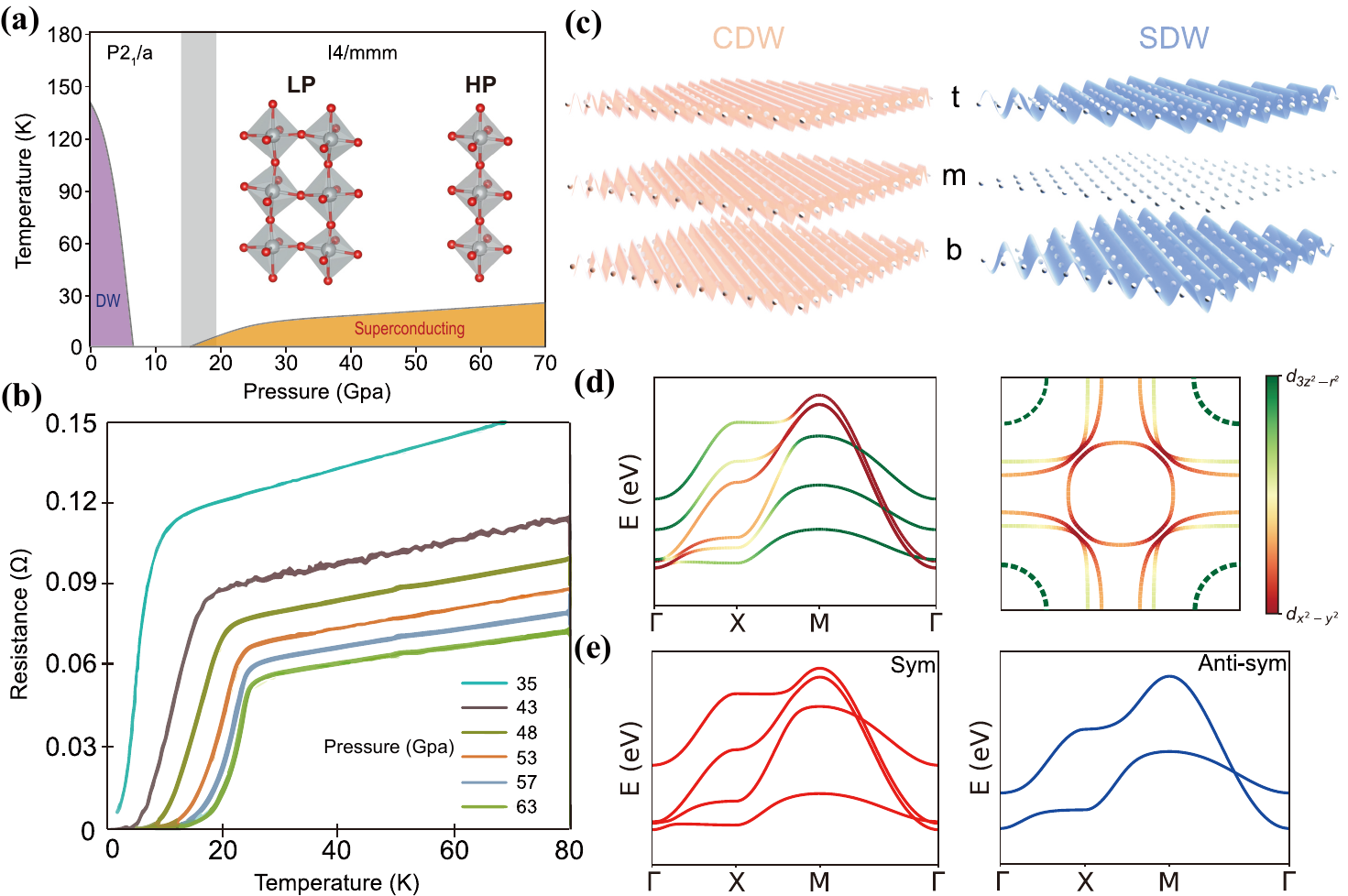}\caption{
        (a) Phase diagram of La$_4$Ni$_3$O$_{10}$ \cite{zhaojun}, featuring both LP and HP phases. The LP phase exhibits a pronounced DW transition, while superconductivity emerges in the HP phase under pressure.
        (b) Resistivity of La$_4$Ni$_3$O$_{10}$ measured under varying pressures, showing a superconducting transition \cite{zhaojun}.
        (c) Density wave phases in the 43(10) compound. The trilayer is commonly labeled as top (t), middle (m), and bottom (b) layers. Notably, the CDW exists in all three layers, whereas the SDW is present only in the top and bottom layers \cite{zhang_intertwined_2020}.
        (d) Band structure and Fermi surfaces of La$_4$Ni$_3$O$_{10}$. The presence of the $d_{3z^2-r^2}$ Fermi surface pocket at the Brillouin zone corner remains unresolved.
        (e) The band structure can also be decomposed into symmetric and antisymmetric sectors, with the symmetric sector resembling that of La$_3$Ni$_2$O$_7$.
			\label{fig10}}
	\end{center}
	\vskip-0.5cm
\end{figure*}

Several ARPES studies have been carried out to investigate the key electronic features underlying superconductivity in 327 thin films \cite{junfeng-arpes1, zxshen-arpes, junfeng-arpes2}. Overall, the observed band structure in these superconducting films resembles that of the LP phase, as shown in Fig. \ref{fig7}(d), where the $\alpha$, $\beta$, and $\gamma$ bands are commonly identified. Among these, the $\gamma$ band—particularly its position relative to the Fermi level—has attracted special attention, given that FS topology is a crucial factor for understanding superconductivity. As shown in Fig. \ref{fig9}, ARPES measurements on La$_2$PrNi$_2$O$_7$ and La$_{2.85}$Pr$_{0.15}$Ni$_2$O$_7$ films reveal distinct FS topologies. In La$_2$PrNi$_2$O$_7$, there is no $\gamma$ pocket observed; its band top lies approximately 70 meV below the Fermi level, even deeper than the 50 meV observed in the LP phase. In contrast, the $\gamma$ band in La$_{2.85}$Pr$_{0.15}$Ni$_2$O$_7$ crosses the Fermi level, forming a small pocket, as plotted in Fig. \ref{fig9}(b,d). These contrasting results likely stem from differences in sample thickness, composition, strain, interfacial reconstruction, or growth conditions. Nonetheless, the absence of a $\gamma$ pocket in superconducting La$_2$PrNi$_2$O$_7$ suggests that the $\gamma$ band is not essential for superconductivity in the 327 system.

\subsection{Theory of 327}

Finally, we want to briefly mention theoretical proposals for high-$T_c$ superconductivity in 327. Generally speaking, there are two approaches for the pairing mechanism of high-$T_c$ superconductors: weak-coupling and strong-coupling. Weak-coupling theories focus on Fermi surface instabilities and typically employ methods such as the random phase approximation \cite{yaodx,yangf,dagotto1,zhang2023trends,PhysRevB.108.L201121,heier2024competing,gu2025effective,xia2025sensitive,xi2025transition,Kuroki,Eremin_PhysRevB.109.L180502} or functional renormalization group \cite{wangqh,jiang2025theory,zhan2025cooperation}. In contrast, strong-coupling theories begin from a Mott insulating or magnetically ordered state driven by strong electron correlations, with superconductivity emerging upon doping, as described by $t-J$ or related models \cite{kun_cpl,taoxiang,zhanggm,wucj,yangyf2,qin2023high,siqm,sugang,tian2024correlation,luo2024high,wang2024self,zheng2025s}. At present, it remains unclear which framework is more appropriate for 327. However, any viable theory must be grounded in experimental observations, particularly the Fermi surface revealed by recent ARPES measurements.

\section{43(10) and other multilayer nickelates}

\begin{figure}
	\begin{center}
		\fig{3.5in}{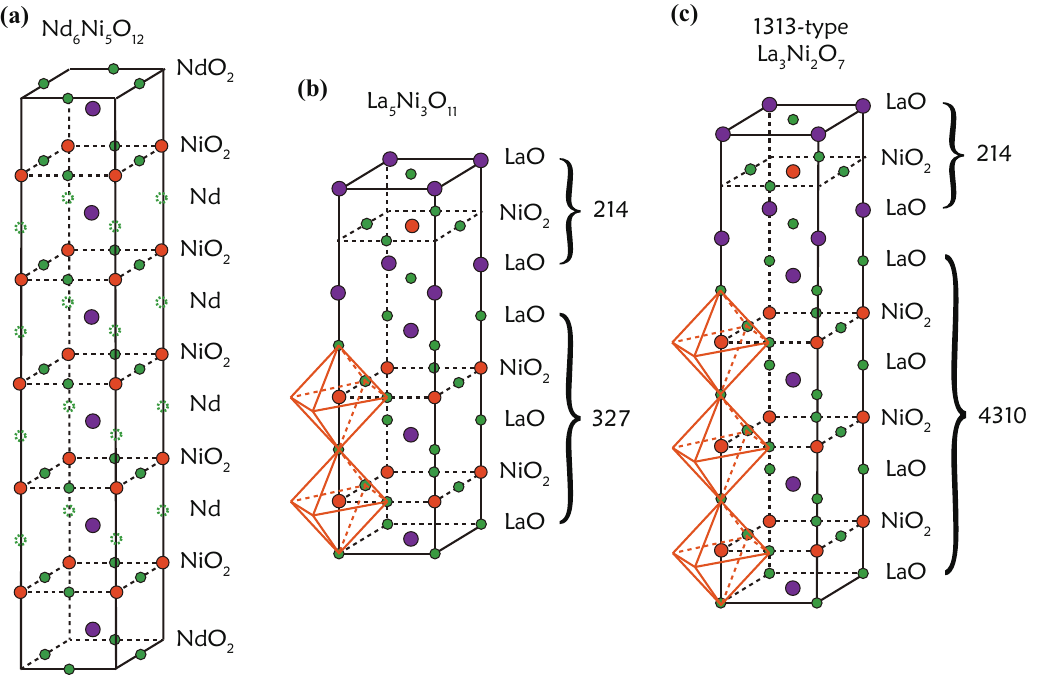}\caption{
        (a) Crystal structure of Nd$_6$Ni$_5$O$_{10}$ \cite{mundy-np}, obtained by reducing the Ruddlesden-Popper phase Nd$_6$Ni$_5$O$_{16}$. (b) Crystal structure of La$_5$Ni$_3$O$_{11}$ \cite{la5ni3o11}, formed by inserting a La$_2$NiO$_4$ layer into the bilayer La$_3$Ni$_2$O$_7$ structure. This compound is also referred to as the 1212 phase. (c) Alternative crystal structure of La$_3$Ni$_2$O$_7$ \cite{Hepting_PhysRevLett.133.146002}, composed of a trilayer La$_4$Ni$_3$O$_7$ unit and a La$_2$NiO$_4$ block. This structure is known as the 1313 phase.
			\label{fig11}}
	\end{center}
	\vskip-0.5cm
\end{figure}

\subsection{43(10)}
The trilayer 43(10) is another case of superconductors under pressure. In the thermodynamic phase diagram, the 43(10) is more stable than 327. The single crystal quality becomes much better than 327. The general phase diagram under pressure and temperature of 43(10) is plotted in Fig. \ref{fig10} \cite{zhaojun}. The phase diagram is also separated into two phases, the high-pressure \textit{I4/mmm} phase and the lower-pressure \textit{P2$_1$/a} phase. Similar to 327, the three NiO$_6$ octahedra line up in the HP structure while they tilt away 180$^{\circ}$ Ni-O-Ni bond angles in the LP structure. 

The transport measurements of 43(10) under pressure in the helium DAC are plotted in Fig. \ref{fig10}(b). After the structural transition around 15 GPa, the HP phase of 43(10) starts to drop in resistance below a critical temperature. The zero resistances are observed above 43.0 GPa. The superconducting $T_c$ onset can reach around 30 K. The d.c. magnetic susceptibility measurements on 43(10) further confirm the diamagnetic response above 30 GPa.

The LP phase undergoes a density wave transition around 136 K at ambient pressure. This density wave transition becomes much sharper than 327. The heat capacity shows an obvious transition around 136 K. The previous single crystal synchrotron x-ray and neutron diffraction have revealed intertwined density waves with a layer-dependent feature in this trilayer structure \cite{zhang_intertwined_2020}. As shown in Fig. \ref{fig10}(c), the SDW displays a node on the inner Ni-O plane while maintaining an out-of-phase between the two outer Ni-O planes. In contrast, the CDW persists across all Ni-O layers with an in-phase \cite{zhang_intertwined_2020}. In addition, the CDW and SDW exhibit incommensurate propagation vectors $\mathbf{q_{c}} = (0,q_{c},0)$ and $\mathbf{q_s} = (0,1-q_{s},0)$,respectively, with $q_{c} = 2q_{s}$ as expected for a system with coupled charge and spin order \cite{zhang_intertwined_2020}. Direct visualization of an incommensurate unidirectional CDW has also been observed by scanning tunneling microscopy (STM) \cite{li_direct_2025}. Complementary experiments, including $\mu$SR \cite{cao_complex_2025,khasanov_identical_2025} and NMR measurements \cite{fukamachi_139nmr_2001,fukamachi_studies_2001,kakoi_multiband_2024}, have independently identified a density wave transition with a transition temperature $T_{\mathrm{DW}}$ $\approx$ 135 K. Furthermore, a distinct formation of a density wave energy gap is revealed by optical conductivity, pump-probe and ARPES measurements \cite{li_fermiology_2017,li_distinct_2025,du_correlated_2024,xu_origin_2025}. Under applied pressure, transition temperatures of both CDW and SDW were suppressed \cite{xu_collapse_2025,khasanov_identical_2025}, which contrasts with the double-layer RP nickelate La$_{3}$Ni$_{2}$O$_{7}$ \cite{nmr,musr2}. Moreover, two $\mu$SR studies report an additional order of SDW at a lower temperature $T_{\mathrm{SDW2}} \approx$ 80 K in both single crystal and polycrystalline La$_{4}$Ni$_{3}$O$_{10}$ samples \cite{khasanov_identical_2025,cao_complex_2025}. However, the underlying mechanism, whether involving a spin reorientation, a phase separation or other phenomena, still remains unclear \cite{khasanov_identical_2025,cao_complex_2025}.

The electronic structure of 43(10) becomes more complicated than 327.
As discussed above, the RP trilayer structure comes from a bilayer 327 with an additional LaNiO$_3$ layer. 
It is better to classify its band structures into two groups, inversion symmetric and inversion antisymmetric.
Using the labeling top (t), middle (m), bottom (b) of the trilayer NiO$_2$ planes, the symmetric bands formed by $\frac{1}{\sqrt{2}}(\psi_t+\psi_b)$ and $\psi_m$ while the antisymmetric bands formed by $\frac{1}{\sqrt{2}}(\psi_t-\psi_b)$.
As shown in Fig. \ref{fig10}(d), the symmetric bands contain four bands, which shows a similar contour of 327.
On the other hand, the antisymmetric bands host two bands like a single-layer two-orbital model.
There are also theoretical groups proposing to further split the symmetric bands. However, due to the on-site energy of the middle is always different from the top and bottom. It is not convenient to treat them separately, especially under electron correlation.
With this information, one can easily understand the FS of 327. Two FSs from symmetric bands like 327, and another additional FS from the antisymmetric band.

\subsection{Other Multilayer Nickelates}
Besides 112, 327, 43(10), there are other multilayer nickelate superconductors or related structures, as plotted in Fig. \ref{fig11}. We briefly list these findings in this short section.

We have shown that LaNiO$_2$ comes from the oxygen reduction of LaNiO$_3$. Therefore, it is natural to achieve other multilayer 112 structures through the reduction of other RP nickelates.  Several multilayer 112 structure thin films have been grown using oxide molecular beam epitaxy \cite{mundy-np}. The crystal structure of Nd$_6$Ni$_5$O$_{10}$ is plot in Fig. \ref{fig11}(a), where oxygens from Nd plane are removed in the RP Nd$_6$Ni$_5$O$_{16}$.  One benefit from this process is that hole-doping naturally occurs. Hence, no chemical doping using Sr or other ions is needed.  The $d^{8.8}$ electron filling is obtained in 5-layer Nd$_6$Ni$_5$O$_{12}$ with onset $T_c$ around 10K. On the other hand, Nd$_4$Ni$_3$O$_8$ with $d^{8.67}$ filling becomes outside the superconducting dome.


As discussed in the introduction, the $n$-layer RP nickelates can be constructed by inserting LaNiO$_3$ layers into the La$_2$NiO$_4$ framework. Conversely, it is also possible to incorporate La$_2$NiO$_4$ layers into RP structures. This concept is realized in the compound La$_5$Ni$_3$O$_{11}$ \cite{la5ni3o11}, which can be viewed as a hybrid of La$_3$Ni$_2$O$_7$ and La$_2$NiO$_4$. In La$_5$Ni$_3$O$_{11}$ shown in Fig. \ref{fig11}(b), single-layer and bilayer NiO$_6$ octahedral blocks alternate along the $c$-axis, forming a unique hybrid RP nickelate known as the 1212 phase. Similar to La$_3$Ni$_2$O$_7$, La$_5$Ni$_3$O$_{11}$ exhibits a density-wave transition near 170 K. However, unlike La$_3$Ni$_2$O$_7$ and La$_4$Ni$_3$O$_{10}$, the transition temperature in La$_5$Ni$_3$O$_{11}$ increases steadily with pressure, reaching approximately 210 K at 12 GPa, before abruptly vanishing prior to the onset of pressure-induced superconductivity at higher pressures. The maximum $T_c$ observed in La$_5$Ni$_3$O$_{11}$ is 64 K, which is slightly lower than that of La$_3$Ni$_2$O$_7$. High-pressure synchrotron X-ray diffraction reveals a structural phase transition in La$_5$Ni$_3$O$_{11}$ from an orthorhombic to a tetragonal symmetry at around 4.5 GPa. Interestingly, unlike in La$_3$Ni$_2$O$_7$ and La$_4$Ni$_3$O$_{10}$, this structural transition does not significantly affect either the density-wave state or the superconducting properties in La$_5$Ni$_3$O$_{11}$.




In addition to La$_3$Ni$_2$O$_7$, recent studies have reported the coexistence of an alternating monolayer–trilayer phase, known as the 1313 phase, alongside the bilayer La$_3$Ni$_2$O$_7$ structure \cite{Hepting_PhysRevLett.133.146002,1313_PhysRevLett.134.126001}. The crystal structure of the 1313 phase is illustrated in Fig. \ref{fig11}(c). Previous high-pressure transport measurements indicate the potential for high-temperature superconductivity in the 1313 phase, with an onset transition temperature of around 80 K. However, considering that the $T_c$ of the trilayer compound La$_4$Ni$_3$O$_{10}$ is only around 30 K, it is likely that the observed superconductivity in the 1313 phase originates from residual bilayer La$_3$Ni$_2$O$_7$, which is inevitably present in the 1313.
The intrinsic physical properties of the 1313 phase, as well as its influence on the superconducting behavior of the 327 system, remain open questions and warrant further detailed experimental and theoretical investigation.

\section{Perspective}
The discovery of superconductivity in nickelates has revitalized interest in the broader field of unconventional superconductors. In this article, we have reviewed the recent significant advances in the study of nickelate superconductors. However, critical challenges and opportunities remain across several key aspects that are shaping the future direction of nickelate research.
\begin{itemize}
	\item \textit{Sample quality}: The synthesis of high-quality nickelate samples—both thin films and single crystals—remains a critical bottleneck in the field. Achieving high-quality samples is essential for definitive conclusions.
    Especially, the 327 sample quality still requires significant improvement. 
	\item \textit{Electronic structure}: Understanding the electronic structure is a fundamental step toward uncovering the origin of superconductivity and related physical phenomena. Recent ARPES studies on 112 have shed light on the interplay between 3d and rare-earth 5d electrons. More ARPES findings on superconducting 327 hold promise for deepening our understanding of the electronic nature that supports superconductivity. 
    \item \textit{Disorder}: Disorder is an unavoidable aspect of nickelate superconductors, with apical oxygen vacancies playing a particularly critical role. Elucidating the impact of such disorder requires further theoretical and experimental investigation.
    \item \textit{Innovative material characterization}: Current material characterization techniques are largely constrained to thin-film samples and high-pressure environments. To advance the field, the development of more innovative and versatile characterization methods is essential.
    \item \textit{Pairing symmetry}: Identifying the pairing symmetry is a crucial step in understanding unconventional superconductors. However, definitive conclusions remain elusive, largely due to limitations in sample quality and the challenges associated with current characterization techniques.
    \item \textit{Difference between cuprates and nickelates}: Insights from the study of nickelates may provide valuable clues for resolving the enduring mysteries surrounding cuprate superconductors. Table \ref{table:1} presents a comparison between these two families. By highlighting both their similarities and differences, we hope this comparison will contribute to the development of a unified framework for understanding high-temperature superconductivity.
 \item \textit{Implication for discovering new high-$T_c$ superconductors}:  The newly discovered nickelate superconductors may also establish important guidelines for searching for new high-$T_{c}$ materials. As shown in Table \ref{table:1}, there exist two distinct electronic environments that can host high-$T_{c}$ superconductivity. These environments are consistent with the recently proposed “genes” framework for unconventional high-$T_{c}$
 superconductors \cite{hujp_PhysRevX.5.041012,jphugene2016}, in which the transition metal $d$-orbitals strongly hybridized with the oxygen $p$-orbitals are isolated near the Fermi energy. With the discovery of new high-$T_{c}$ superconductors, summarizing their common features to guide the search for other novel high-$T_{c}$ materials is also a worthwhile direction of exploration.
\end{itemize}

\begin{widetext}


\begin{table}[htbp]
\centering
\begin{tabular}{|l|c|c|c|c|}
\hline
 & \textbf{Cuprates} & \textbf{LaNiO\textsubscript{2}} & \textbf{La\textsubscript{3}Ni\textsubscript{2}O\textsubscript{7}} & \textbf{La\textsubscript{4}Ni\textsubscript{3}O\textsubscript{10}} \\
\hline
Layered Structures & 2D CuO\textsubscript{2} planes & 2D NiO\textsubscript{2} planes & Bilayer NiO\textsubscript{2} planes & Trilayer NiO\textsubscript{2} planes \\
\hline
Valence State & Cu\textsuperscript{2+} (3d\textsuperscript{9}) & Ni\textsuperscript{1+} (3d\textsuperscript{9}) & Ni\textsuperscript{2.5+} (3d\textsuperscript{7.5}) & Ni\textsuperscript{2.67+} (3d\textsuperscript{7.33}) \\
\hline
Orbital Character & 3d\textsubscript{$x^2$–$y^2$} & \makecell{3d\textsubscript{$x^2$–$y^2$} \\ hybridized with rare-earth orbitals} & 3d\textsubscript{$x^2$–$y^2$}, 3d\textsubscript{$3z^2-r^2$} & 3d\textsubscript{$x^2$–$y^2$}, 3d\textsubscript{$3z^2-r^2$} \\
\hline
Parent State & Charge Transfer insulator & \makecell{Mott insulator \\(without rare-earth)} & \makecell{Open question} & \makecell{Open question} \\
\hline
Superconductivity & $d$-wave SC & Evidence of $d$-wave SC & \makecell{pressure or thin film, \\ symmetry unknown} & \makecell{pressure, \\ symmetry unknown}\\
\hline
$T_c$ & $\sim$135 K & $\sim$15–40 K & \makecell{$\sim$80 K (high pressure) \\ 48 K (thin film)} & $\sim$40 K (high pressure) \\
\hline
Correlation Strength & Strong & Strong & Strong or intermediate & Strong or intermediate \\
\hline
Electron-Phonon for SC & $\times$ & Unlikely & Unlikely & Unlikely \\
\hline
\end{tabular}

\caption{Comparison between Cuprate, LaNiO\textsubscript{2}, La\textsubscript{3}Ni\textsubscript{2}O\textsubscript{7}, and La\textsubscript{4}Ni\textsubscript{3}O\textsubscript{10} superconductors.}

\label{table:1}
\end{table}

\end{widetext}

\section{Acknowledge}
We thank Fu-Chun Zhang, Tao Xiang, Donglai Feng, Meng Wang, Guangming Zhang, Qianghua Wang, Yijun Yu, Danfeng Li, Huiqiu Yuan, Xingjiang Zhou, Zhen Chen, Ke-Jin Zhou, Dawei Shen, Junfeng He, Yuefeng Nie, Lei Shu, Jun Zhao, Ariando Ariando,  Bai Yang Wang, Kyle Shen, Berit Goodge et. al. for useful discussions. 

K.J. and J.P.H. acknowledge the support of the National Natural Science Foundation of China (Grant NSFC-12494594,  No. NSFC-12174428), the New Cornerstone Investigator Program. K.J., J.J.Y., and W.T. acknowledge the Chinese Academy of Sciences Project for Young Scientists in Basic Research (2022YSBR-048).  J.J.Y. acknowledges support from the National Natural Science Foundation of China (Grant NSFC-12494592). W.T. acknowledges the National Natural Science Foundation of China (Grant No. 12325403). W.T. and X.H.C. acknowledge the National Key R\&D Program of the MOST of China (Grant 2022YFA1602601). W.T., J.J.Y., and X.H.C. acknowledge the Chinese Academy of Sciences under contract No. JZHKYPT-2021-08. J.G.C. acknowledges support from the National Natural Science Foundation of China (Grant NSFC-12025408, U23A6003), and the National Key R\&D Program of China (2023YFA1406100).

\bibliography{ref}

\end{document}